\documentclass[aps,pra,epsfigure,twocolumn]{revtex4}
\usepackage{amsmath}
\usepackage{amstext}
\usepackage{latexsym}
\usepackage{graphicx}
\usepackage{amsfonts}
\usepackage{epsfig}
\usepackage{color}

\newcommand{\Tr}{\mathrm{Tr}}

\begin{document}
\title{Optimal probabilistic measurement of phase}
\author{Petr Marek}
\affiliation{Department of Optics, Palack\' y University,\\
17. listopadu 1192/12,  771~46 Olomouc, \\ Czech Republic}
\date{\today}
\begin{abstract}
When measuring phase of quantum states of light, the optimal single-shot measurement implements projection on the un-physical phase states. If we want to improve the precision further we need to accept a reduced probability of success, either by implementing a probabilistic measurement or by probabilistically manipulating the measured quantum state by means of noiseless amplification. We analyze the limits of this approach by finding the optimal probabilistic measurement which, for a given rate of success, maximizes the precision with which the phase can be measured.
\end{abstract}
\maketitle

Phase is a central concept in both classical and quantum optics. It was, however, a matter of lengthy dialogue, before the quantum description of phase was established. The initial attempts of Dirac to treat phase as a canonical conjugate to photon number failed, because it is impossible to represent phase by a quantum mechanical observable \cite{phase_old}. As a consequence, phase can not be projectively measured, it can only be estimated (or guessed) by analyzing the results of other measurements. Despite this, phase states do exist \cite{phase_state} (even if they are not orthogonal) and they were eventually used to construct a well behaved phase operator \cite{phase_operator}. Other attempts to describe phase properties of quantum states relied on the measurement-related phase distribution \cite{phase_measurement}. Both approaches were later reconciled with the fundamental canonical phase distribution \cite{phase_canonical}.

The canonical phase distribution characterizes phase properties of a quantum state and it is completely independent of its photon number distribution. It can be used to obtain a wide range of quantities related to phase estimation, but it also determines how much information about the phase of the state can be obtained by performing a measurement only on a single copy of it. True, the ideal canonical phase measurement does not and cannot exist, but several approximative approaches have been suggested \cite{canonical_measurement1,canonical_measurement2}.

Aside from improving the actual detector scheme, overall performance of phase measurement can be also enhanced by specific alteration of the measured quantum state. Highly nonclassical quantum state can, in principle, lead to an unparalleled precision \cite{Noon}, while weakly nonclassical states are both beneficial and experimentally feasible \cite{phase_tracking}. However, if the state is unaccessible prior to phase encoding, we need to rely on operations which can enhance the amount of phase information already carried by the scrutinized state. Such operations are commonly referred to as noiseless amplifiers and a great deal of attention was recently devoted both to the concept \cite{noiseless1} and to the experimental realizations \cite{noiseless2}. The cost of this improvement comes in the reduced success rate of the operation. The amplification is therefore not very practical when the measurements can be repeated, but ut may be useful when the event to be detected is rare and we need to be certain that the single obtained measurement outcome corresponds to the theoretical value as closely as possible.

However, even in the scenarios in which the probabilistic approach is worth considering, it would be more prudent to design an actual probabilistic measurement of phase. Such the measurement would be conceptually similar to methods of unambiguous discrimination of quantum states \cite{discrimination}, except that a truly error-less detection would be possible only in the limit of zero probability. Rather then this regime of limited interest, the question is: how does reducing the success rate of the measurement help us to measure the phase more precisely. And, maybe even more importantly, what are the theoretical limits of this approach?  In this paper we attempt to answer these questions.

Let us start by reviewing what we actually mean by the term `phase measurement'. Phase has well defined meaning only in the context of an interferometric setup, where it expresses the relative length difference between the two optical paths. In the context of continuous variables (CV) quantum optics \cite{CVoptics}, phase is often considered a stand-alone property.  However, this is only because the other path in the interferometer, represented by the local oscillator, is taken for granted. In a sense this is justified, as the local oscillator is intense enough to be, for all intents and purposes, just a classical reference framing the associated quantum system. Measuring the phase of the quantum system is then equivalent to discerning a value of parameter $\phi$, which was encoded into the quantum state by means of an operator $\exp( i \phi \hat{n})$, where $\hat{n}$ is the photon number operator. Apart from special cases it is impossible to determine the parameter $\phi$ perfectly. Rather than complete knowledge, the result of the measurement provides the observer just with the best guess of the parameter, where the quality of the guess depends on both the state of the measured system and the phase measurement employed.

The simplest single-shot measurement of phase of optical signals relies on simultaneous measurement of quadrature operators $X$ and $P$, corresponding to the Hermitian and the anti-hermitian part of annihilation operator. The phase can be then deduced from the measurement results $x'$ and $p'$ by taking $\phi = \tan^{-1}(p'/x')$. Of course, in addition to knowledge of phase, this particular measurement also provides us with knowledge of the energy of the state. Therefore the obtained phase information is not as complete as it could be.

The best possible measurement which can be imagined is the so called canonical measurement of phase. It can be mathematically described as a projection on idealized phase states $|\theta\rangle = \sum_{k=0}^{\infty} e^{i\theta k}|k\rangle$. These phase states are not normalized, which makes them similar to eigenstates of continuous operators (such as position and momentum), but they are also not orthogonal. The non-orthogonality is actually responsible for the impossibility to measure phase completely, because a single measured value of $\theta$ is not exclusive just to a single phase state. For any quantum state $\hat{\rho}$ the results of the canonical phase measurement can be characterized by probability distribution $P(\theta) = \Tr [\hat{\rho} |\theta\rangle\langle \theta|]$ - the canonical phase distribution. The shape of the distribution is solely given by the employed quantum state, the encoded phase value is represented only as a linear displacement. For a particular measured value $\theta$ the value $|P(\theta)|$ is related to the probability that the measured value is the encoded value.  Simplistically, we can say that for any quantum state, the quality of phase encoding is given by the width of the canonical distribution. This can be formally done by taking the variance of the phase distribution, but it is actually more convenient to use a different quantity. One, which takes into account the periodicity of the phase on interval $\langle 0,2\pi\rangle$ \cite{dispersion}. The new quantity is the phase variance $V =  |\mu|^{-2} -1 $, where $\mu = \langle \exp{i\theta} \rangle $ \cite{Helstrom}. The phase variance is completely independent of displacement in $\theta$, it is therefore completely determined by the state $\hat{\rho}$. We can also see that the phase variance solely depends on the value of parameter $\mu$, which we are going to use from now on.

\begin{figure}
\begin{center}
\includegraphics[width= 0.95\linewidth]{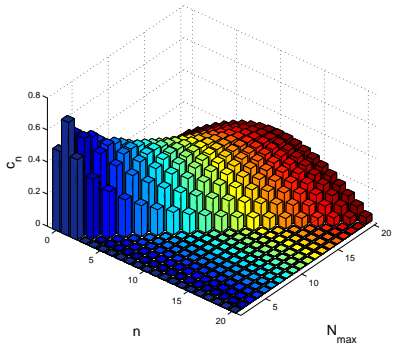}
\end{center}
\caption{(color online). Fock representation of optimal phase states for various sizes of Hilbert space (represented by value of $N_{max}$. )  }
\label{fig_optimalstate}
\end{figure}

It is instructive to look at phase properties of physical quantum states and find out, which states are best suited for encoding of phase. And while our main interest lies in states from infinite dimensional Hilbert space, it is practical to start by limiting ourselves to a Hilbert space with a finite dimension $N$. In this limited Hilbert space with basis states $|n\rangle$, any quantum state can be expressed as a superposition
\begin{equation}\label{}
    \sum_{n=0}^N c_n |n\rangle, \quad \mathrm{with}~\sum_{n= 0}^{N} |c_n|^2 = 1.
\end{equation}
For this state, the modus of the parameter $\mu$, which is the sole factor responsible for the phase variance, can be obtained as
\begin{equation}\label{mu1}
    \mu = |\sum_{n=0}^{N-1} c_n c_{n+1}^*|.
\end{equation}
The optimal state for phase encoding - the state which leads to minimal phase variance - can now be obtained by maximizing $\mu$ (\ref{mu1}) under the condition $\sum_{n=0}^{N} c_n^2 = 1$. We can start by observing that for any particular values of $|c_n|$ the maximum will be obtained when
\begin{equation}\label{}
     \frac{c_{n+1}}{c_n} = r_n e^{i\varphi}, \quad \forall n = 0,\ldots,N-1
\end{equation}
where $\{r_n\}$ is a sequence of positive real numbers and $\varphi$ is a real number same for all the pairs of coefficients. With no loss of generality we can therefore set value of $\varphi$ to zero and in the following consider only quantum states which have all their coefficients $c_n$ real and positive. Using Lagrange multipliers, we can find recursive relations for the coefficients:
\begin{equation}\label{}
    c_{n+1} = \lambda c_n - c_{n-1},
\end{equation}
with $c_{-1} = c_{N+1} = 0$ by convention. This allows us to express any coefficient with help of a polynomial of $\lambda$ as:
\begin{equation}\label{}
    c_n = \mathcal{P}_n (\lambda) c_0.
\end{equation}
We can now take advantage of the symmetry of (\ref{mu1}), which ensures that $c_N = c_0$. We can then find the appropriate value of $\lambda$ as the real and positive root of polynomial equation $\mathcal{P}_N(\lambda)=1$, which provides us with the highest value of (\ref{mu1}). The value of $c_0$ in the formula is simply given by normalization $c_0 = [\sum_{n=0}^N \mathcal{P}_n^2(\lambda)]^{-1}$. Optimal states for varying size of the Hilbert state are depicted in Fig.~\ref{fig_optimalstate}.

Existence of the ideal state tells us there are limits to how well can the phase be encoded in a limited-dimensional Hilbert space. On the other hand, if the Hilbert space is infinite, which is the case in CV quantum optics communication, it is in principle possible to encode the phase perfectly - in such the way that $\mu = 1$ and consequently the phase variance is zero. As this is obviously the case in classical communication, where phase can be encoded and decoded with arbitrary precision, the inability to measure phase in quantum physics stems from employing quantum states which are so weak their Hilbert space is effectively limited. However, there is a key difference between these states and states from a Hilbert space with factually limited dimension. The difference being that the infinite dimensional Hilbert space offers a possibility of measuring the state arbitrarily well if we accept reduced probability of success.

The idea that measurement can be improved when we accept a reduced probability of success is not a new one. When discriminating quantum states drawn from a finite ensemble, one can accept existence of inconclusive results (reduced success rate) in order to reduce the probability of erroneous result to zero \cite{discrimination}.  Similarly, when measuring a continuous parameter such as phase, it is possible to conditionally transform the quantum states in such the way that the subsequent measurement leads to more precise results \cite{noiseless1, noiseless2}. Taken as whole, the combination of probabilistic operation and measurement is essentially a probabilistic measurement. In the following we develop a unified picture describing probabilistic measurement of phase of a quantum state and derive bounds for the optimal one. Namely we will look for such the measurement, which for a given probability of success yields the best possible result.

Extension of the canonical measurement of phase into the probabilistic regime can be represented by a set of operators $\Pi_{\phi}$, each of them corresponding to a positive detection event of a value $\phi$, and a single operator $\Pi_0$ representing the inconclusive results. Together these operators form a positive operator valued measure (POVM). For the canonical deterministic measurement of phase these operators are $\Pi^{(D)}_{\phi} = \frac{1}{2\pi}|\phi\rangle\langle\phi|$. Keeping the pure-state projector structure intact, we can express the probabilistic POVMs as
\begin{equation}\label{}
    \Pi^{(P)}_\phi = \frac{1}{2\pi} F|\phi\rangle\langle\phi|F^{\dag}, \quad \Pi^{(P)}_0 = 1 - \int \Pi^{(P)}_\phi d \phi.
\end{equation}
Here $F = \mathrm{diag} (f_0,f_1,\cdots)$, where $|f_j|\leq 1$ for all $j = 0,1,\cdot$, is operator diagonal in Fock space. It is practical to represent the probabilistic measurement by a filter, transmitting and modifying the quantum state with some limited probability, followed by the deterministic canonical phase measurement. The operator $F$ then plays the role of the probabilistic filter and the task of finding the optimal measurement is reduced to finding the optimal operator $F$.

After the first glance at the problem, one issue immediately becomes apparent. For any quantum state $\rho$, the probability of successful measurement, $P = 1 - \Tr[\rho \Pi^{(P)}_0]$ is dependant on the choice of the measured state. The optimal measurement therefore needs to be tailored to a specific state or to a class of states. But let us first approach the task in the general way. Suppose we have an input quantum state
\begin{equation}\label{}
    |\psi\rangle = \sum_{n=0}^{\infty} c_n |n\rangle.
\end{equation}
We have previously shown that it is best for phase encoding when $c_n>0$ for all $n$, so we will assume this is the case \cite{comment}.  The act of the filter transforms this state into a new one,
\begin{equation}\label{}
    |\psi_f\rangle = \frac{1}{\sqrt{P}}\sum_{n=0}^{\infty} f_n c_n |n\rangle,
\end{equation}
where $P = \sum_{n=0}^{\infty}f_n^2 c_n^2$ is the probability of success and the filter parameters $f_n$ were also considered real and positive. For any given probability $P$, the act of finding the optimal filter can be reduced to solving a system of equations
\begin{eqnarray}\label{}
    f_{n-1} a_{n-1} + f_{n+1} a_n = \lambda f_n x_n, \quad n = 0,1,\ldots, \nonumber \\
    \sum_{n=0}^{\infty} x_k f_k^2 = P,
\end{eqnarray}
where $a_n = c_n c_{n+1}$, $x_n = c_n^2$, $f_{-1} = 0$ by convention, and $\lambda$ is Lagrange multiplier. Finding the solution under the most general conditions is not an easy task. Fortunately, there are some simplifications which can be made, provided we are applying the filtration to the practically significant coherent states.

\begin{figure}
\begin{center}
\includegraphics[width= 0.95\linewidth]{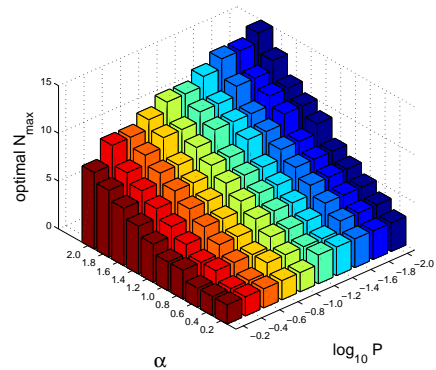}
\end{center}
\caption{(color online). Optimal filter parameters $N$ in dependance on coherent amplitude of the coherent state $\alpha$ and the probability of the successful measurement $P$. }
\label{fig_optimalN}
\end{figure}
\begin{figure}
\begin{center}
\includegraphics[width= 0.8\linewidth]{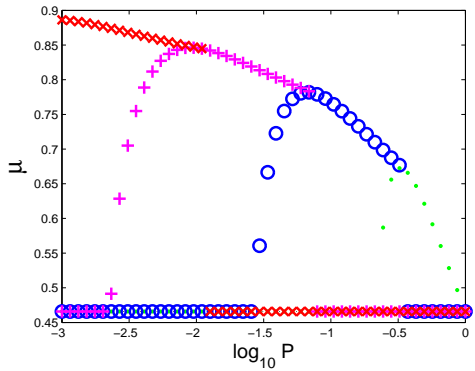}
\end{center}
\caption{(color online). Value of $\mu$ for optimal probabilistic measurement of phase of coherent state with $\alpha = 0.5$ in dependance on probability of success. Different colors and markers denote values for particular choices of $N$: $N=1$ (green dot), $N=2$ (blue circle), $N=3$ (magenta '+' cross), and $N=4$ (red 'x' cross). When the particular choice of $N$ does not yield optimal physical filter for a particular value of $P$, the value of $\mu$ displayed is the initial value of deterministic measurement.}
\label{fig_muforalpha0_5}
\end{figure}

A coherent state $|\alpha\rangle = \sum_{k=1}^{\infty} \frac{\alpha^k}{\sqrt{k!}}|k\rangle$ can be considered a quantum version of a classical complex amplitude of light. It can be used to describe the state of light produced by a well stabilized laser and it has in place both in the classical communication \cite{coh_communication} and in quantum cryptography \cite{coh_cryptography}, both of which can employ phase encoding. Coherent states are fairly well localized in the Fock space - for any coherent state there always exists a finite $N$-dimensional Fock subspace such that the probability of the state manifesting outside of it can be made arbitrarily small. As a consequence, those higher Fock dimensions do not significantly contribute to the state's properties and the values of the respective filters can be set to one, i.e. $f_n =1$ for all $n\geq N$. Of course, with severe filtering leading to extremely low success rates, some previously dismissable Fock numbers can start being relevant, but this can be remedied by choosing even higher photon number $N'$ as the new threshold of significance.

This dramatically simplifies the process of finding the optimal filter. All the filter coefficients for $n = 0,\cdots,N$ can be now expressed in the form
\begin{equation}\label{}
    f_n = f_0 \mathcal{P}_n (\lambda),
\end{equation}
where $\mathcal{P}_n(\lambda)$ is polynomial of $\lambda$ defined by the recursive relation
\begin{equation}\label{}
    P_{n+1}(\lambda) = \frac{\lambda x_n \mathcal{P}_n(\lambda)- a_{n-1} \mathcal{P}_{n-1}(\lambda)}{a_n}
\end{equation}
with $\mathcal{P}_0(\lambda)\equiv 1$ and $\mathcal{P}_1(\lambda) = x_0/a_0$. Since $f_0$ can be obtained from the condition $f_N = f_0\mathcal{P}_N(\lambda) = 1$, getting the full solution is reduced to finding the correct value of lagrange multiplier $\lambda$, which is one of the roots of polynomial equation
\begin{equation}\label{}
    \sum_{n = 0}^N \mathcal{P}_n(\lambda)^2 - \left(P - 1 + \sum_{n=0}^N x_n \right)\mathcal{P}_N(\lambda)^2.
\end{equation}
To be of physical relevance, the obtained $\lambda$ needs to be real and it has to lead to a filter with parameters, which are all positive and bounded by one. And among the values of $\lambda$ satisfying those condition, the one corresponding to the global extreme, rather than just a local one, needs to be selected by directly checking the respective value of $\mu$.

This approach yields the optimal filter for arbitrary coherent amplitude $\alpha$ and arbitrary probability of success $P$. But for any such combination, there is only one choice of $N$ for which it does so. The particular choice of $N$ needs to be found numerically, but that is a simple matter of checking a range of values of $N$ and finding the one which leads to positive results. For illustration, several values of $N$ optimal for some range of $\alpha$ and $P$ are depicted in Fig.~\ref{fig_optimalN}. As another illustration, Fig.~\ref{fig_muforalpha0_5} shows quality of the probabilistic measurement, represented by parameter $\mu$, in dependance on the probability of success. We can see that a specific choice of $N$ yields optimal filter only for a limited range of probabilities, but that for any probability there exists one.

We have introduced the concept of optimal probabilistic measurement of quantum phase and shown how such the measurement can be constructed. The approach can be used for any quantum state, but we have mainly focused on practically relevant coherent states, for which we have managed to obtain the form of the optimal measurement in a semi-analytic form. The probabilistic aspect of the measurement can be represented by a filter transmitting various Fock space elements with different amplitudes. The derived optimal measurement sets an upper bound on the trade-off between the quality and the probability of success of phase measurements. The filter required for such the measurement is a highly non-linear operation, but in light of the recent advent of manipulating light on the individual photon level \cite{noiseless2}, it might be within the experimental reach.

\textbf{Acknowledgements.}
We would like to thank Michal Mi\v{c}uda and Zdenek Hradil for valuable and fruitful discussions. The work was supported by project P205/12/0577 of Grant Agency of Czech Republic.


\begin{thebibliography}{99}
\bibitem{comment}
If this was not the case, the filter could be used to remedy that by adding suitable phase factors and improve the phase properties of the state at no cost in the success rate.
\end{thebibliography}
\end{document}